\documentclass[aps,prl,twocolumn,showpacs,floatfix,superscriptaddress]{revtex4}
\usepackage{graphicx}
\usepackage{bm}
\usepackage{amsmath}
\begin{document}
\title{Multiparticle Interference, 
GHZ Entanglement, and Full Counting Statistics}
\author{H.-S. Sim}
\affiliation{
Department of Physics, Korea Advanced Institute of Science and Technology,
Daejeon 305-701, Korea
}
\author{E. V. Sukhorukov}
\affiliation{
D\'{e}partement de Physique Th\'{e}orique, Universit\'{e} de Gen\`{e}ve,
CH-1211 Gen\`{e}ve 4, Switzerland}
\date{Received 13 August 2005}
\begin{abstract}
We investigate the quantum transport in a generalized $N$-particle 
Hanbury Brown--Twiss setup enclosing magnetic flux, and demonstrate
that the $N$th-order cumulant of current cross correlations
exhibits Aharonov-Bohm oscillations, while there is no such oscillation 
in all the lower-order cumulants. The multiparticle interference results 
from the orbital Greenberger-Horne-Zeilinger entanglement of $N$ 
indistinguishable particles. For sufficiently strong Aharonov-Bohm 
oscillations the generalized Bell inequalities may be violated,
proving the $N$-particle quantum nonlocality.

\end{abstract}


\maketitle

The Aharonov-Bohm (AB) effect \cite{Aharonov1}, 
being a most remarkable manifestation of quantum coherence,
is at the heart of quantum mechanics. 
It is essentially a
{\em topological} effect, because it requires a multiple-connected
physical system, e.g.\ a quantum ring, and consists in 
a periodic variation of physical observables as a
function of the magnetic field threading the loop. 
It is also {\em nonlocal} effect, since no local physical observable 
is sensitive to the field. Originally introduced for a single particle 
\cite{Aharonov1}, it can be generalized as a two-particle AB effect     
in the average current \cite{Loss}, or in the noise power \cite{Buttiker}, 
if only two particles are able to enclose the loop. 

The last effect \cite{Buttiker}, being implemented in the  
Hanbury Brown and Twiss geometry (HBT) \cite{Hanbury},
shows the two-particle character 
in a most dramatic way \cite{Samuelsson}, 
because single-particle observables, such as the average current, 
do not contain AB oscillations. 
In contrast to the single-particle 
AB effect, which may have a classical analog \cite{Aharonov2}, 
the two-particle 
AB effect is essentially quantum, because it originates from quantum 
indistinguishability of particles. Moreover, it is strongly related to the 
orbital {\em entanglement} in HBT setup \cite{Samuelsson}, and to the 
{\em quantum nonlocality} as expressed via the violation of Bell inequalities 
\cite{Bell,CHSH}. 

\begin{figure}[hb]
\includegraphics[width=0.45\textwidth,height=0.2\textheight]{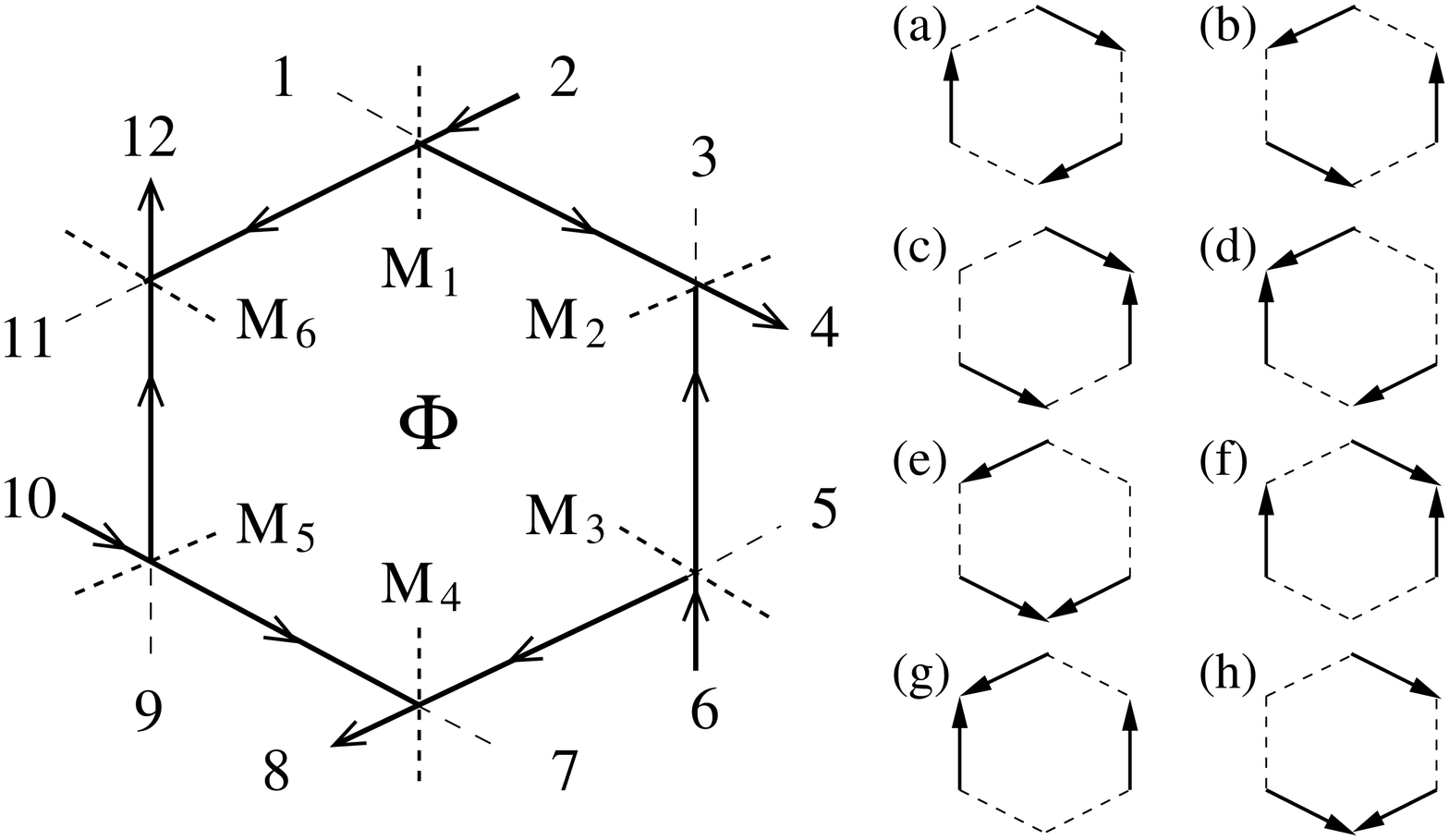}
\caption{Left: Hanbury Brown--Twiss (HBT) interferometer with
$N=3$ independent electron sources (reservoirs 2, 6, 10), 
$2N$ beam splitters $M_i$,  
$2N$ chiral current paths, each connecting two beam splitters, 
and $2N$ detector reservoirs, 3, 4, 7, 8, 11, and 12.
Arrows indicate electron trajectories.
In the setup, any single electron can not enclose magnetic flux $\Phi$,
while $N$ electrons emitted from $N$ sources can do so.
For example, electrons incident from the reservoir 2 move towards
the splitters $M_6$ or $M_2$ and disappear in the detectors
3, 4, 11, or 12. 
This HBT setup can be easily generalized to the case of arbitrary  
$N$.
Right: In the HBT interferometer
$N$-electron states incident from the sources
are decomposed into $2^N$ states, shown for $N=3$.
The states (a) and (b) result in Aharonov-Bohm (AB) oscillations
in the $N$th-order cumulant of the current cross-correlations
at detectors. These two states together form 
a Greenberger-Horne-Zeilinger-type (GHZ) state of $N$ pseudo-spins.
All the others, (c)-(h), do not contribute to the AB effect.}
\label{geometry}
\end{figure}

In this Letter, we investigate the generalized HBT setup,
exemplified in Fig.\ \ref{geometry}, and introduce 
the $N$-particle AB effect for $N\geq 3$. 
We start with analyzing the full counting statistics (FCS) 
\cite{Levitov,Nazarov} and demonstrating that in the $N$-particle 
HBT setup the $N$th-order cumulant of current cross-correlations 
may exhibit AB oscillations, while there is no such oscillation 
in all the lower-order cumulants. 
Next we show that the $N$-particle AB effect originates from the 
orbital $N$-particle entanglement. Namely, we prove that a 
many-particle state injected into the HBT setup contains the 
Greenberger-Horne-Zeilinger-type (GHZ) state (\ref{GHZ}) \cite{GHZ}, which via the 
{\em postselection} \cite{Shih} contributes to the particle transport. 
We remark here that the concept of the FCS has been used in earlier  
works in the context of a two-  
\cite{two} and three-particle \cite{three} entanglement.
We further demonstrate that the Svetlichny inequality 
\cite{Svetlichny} may be formulated in terms of the joint detection probability
and then violated, if the visibility of the AB oscillations 
exceeds the value $1/\sqrt{2}$. This inequality, being most restrictive
in the whole family of generalized $N$-particle Bell inequalities
\cite{Svetlichny,MABK,ZB,Chen},
discriminates quantum mechanics and all hybrid local-nonlocal theories \cite{Gisin}, 
proving the $N$-particle quantum nonlocality in the generalized HBT setup.
We note that in contrast to the two-particle case, the 
overall sign of the $N$th-order cumulant cannot be uniquely associated 
with statistics of particles in the cases of $N \ge 3$. 

{\it Generalized HBT interferometer.}---
We consider the coherent electron transport in the mesoscopic multiterminal
conductor consisting of $4N$ electron reservoirs, $i=1,2,\ldots,4N$
(enumerated clockwise), $2N$ beam splitters $M_i$, $i=1,2,\cdots,2N$, 
and $2N$ chiral current paths enclosing magnetic flux $\Phi$ 
(see Fig.\ \ref{geometry}).
The $(2i-1)$-th and the $2i$-th reservoirs
couple to the
splitter $M_i$, 
at which
electrons can be transmitted with probability $T_i$
or reflected with probability 
$R_i= 1 - T_i$.
Among $4N$ reservoirs, $N$ reservoirs,
$2, 6, \cdots, 4N-2$, behave as {\em independent} electron sources,
and $N$ pairs of reservoirs with indexes 
$\alpha_1=\{3,4\}$, $\alpha_2=\{7,8\}$, $\cdots$, $\alpha_N=\{4N-1,4N \}$,
act as detectors. All sources are biased with voltage $eV$, while 
the rest of $3N$ reservoirs are grounded.
Neighboring beam splitters $M_{i}$ and $M_{i+1}$ are 
connected via a one-dimensional spinless current path, 
where phase difference $\phi_i$ is accumulated.

The total scattering matrix of the generalized HBT setup
consists of two $2N$-dimensional unitary chiral blocks 
$\hat U$ and $\hat V$. 
The block $\hat U$ is determined as follows: 
The electron propagation 
from source $4i+2$ to detectors $4i-1$, $4i$, $4i+3$, or $4i+4$ gives 
the matrix elements
$\exp(- i \phi_{2i}) \sqrt{T_{2i+1} T_{2i}}$,
$i\exp(- i \phi_{2i}) \sqrt{T_{2i+1} R_{2i}}$,
$-\exp{(i \phi_{2i+1})} \sqrt{R_{2i+1} R_{2i+2}}$,
$i\exp{(i \phi_{2i+1})} \sqrt{R_{2i+1} T_{2i+2}}$,
respectively. 
The block $\hat V$ describes noiseless outgoing currents 
from detector reservoirs. It is not relevant  
for the following discussion, therefore the requirement of the 
coherence in this sector may be relaxed. 
All the other matrix elements are zero, 
thus no single electron can enclose the flux $\Phi$ 
in the generalized HBT setup. 
However, $N$ electrons can do so, and this leads 
to the $N$-particle interference in the FCS.

{\it N-particle Aharonov-Bohm effect.}--- 
In the long measurement time limit $t\gg\tau_C$, 
where $\tau_C = 2\pi\hbar/|eV|$ is the correlation time, 
the electron transport is a Markovian random
process. The characteristic property of such a process 
is that the irreducible correlators (cumulants) of the number of
electrons arriving at detectors are proportional 
to the number of transmission attempts, $t/\tau_C$.
Therefore, we normalize the cumulants to $t/\tau_C$, 
so that the cumulant generating function of the FCS at the 
detector reservoirs takes the form 
\cite{Levitov}: 
\begin{eqnarray}
S(\chi) = (\tau_C/2\pi \hbar) \int dE\,
{\rm Tr}\ln [1\!\!1 - \hat{f} +
\hat{f} \hat{U}^\dagger \hat{\Lambda} \hat{U} 
\hat{\Lambda}^\dagger].
\label{Generator}
\end{eqnarray}
Here $\hat{f}$ is the diagonal matrix with elements 
$\hat{f}_{ij} = \delta_{ij} f_i (E)$
being Fermi-Dirac occupations: 
$f_i(E)=f_F(E-eV)$ at the sources, and 
$f_i(E)=f_F(E)$ in the rest of reservoirs.
The matrix $\hat{\Lambda}$ is diagonal with elements
$\hat{\Lambda}_{ij} = \delta_{ij} \exp{(i \chi_i)}$,
where $\chi=\{\chi_i\}$ is the set of counting variables in 
the detector reservoirs. The cumulants may be obtained by evaluating
the derivatives of $S(\chi)$ and setting $\chi=0$.

The generating fuction (\ref{Generator}) is simplified
by introducing
matrices $(\hat A_{\alpha_i})_{nm} =  f_{n}U^*_{\alpha_i,n}U_{\alpha_i,m}$
\cite{footnote1}:
\begin{equation}
S(\chi) = \frac{\tau_C}{2\pi \hbar} \int dE\,
{\rm Tr}\ln\bigg[1\!\!1 +\sum_{\alpha_i}(e^{i\chi_{\alpha_i}}-1)\hat A_{\alpha_i}\bigg],
\label{Generator-A}
\end{equation}
where the sum runs over all the detectors $\alpha_i$ and
$n,m = 1,2,5,6,\ldots,4N-3,4N-2$.
In order to generalize the two-particle interference to the cases of $N\geq 3$, 
we consider the lowest-order cross-correlation functions with all detectors
being different. 
It turns out that the number of such cross-correlators
is limited, because between all the possible lowest-order products
of the form $\prod_{\alpha_i}\hat A_{\alpha_i}$, where all $i$'s are different,
only few products give nonvanishing traces:
${\rm Tr}\hat A_{\alpha_i}$, ${\rm Tr}\hat A_{\alpha_i}
\hat A_{\alpha_{i+1}}$ for all $i$, and 
${\rm Tr}\hat A_{\alpha_1}\hat A_{\alpha_2}\cdots\hat A_{\alpha_N}$ and 
${\rm Tr}\hat A_{\alpha_N}\hat A_{\alpha_{N-1}}\cdots\hat A_{\alpha_1}$.
Expanding the logarithm in the r.h.s.\ of Eq. (\ref{Generator-A})
and collecting nonzero traces we obtain the first-order cumulant
$Q_{\alpha_i}=\tau_CI_{\alpha_i}$ (i.e.\ the normalized average current)
and the second-order cumulant $Q_{\alpha_i,\alpha_{i+1}}$
(the normalized zero-frequency noise power), 
\begin{subequations}
\label{cumulants}
\begin{eqnarray}
Q_{\alpha_i} & = & P_{\alpha_i} R_{2i-1} + (1-P_{\alpha_i})T_{2i+1}
\label{firstCUMULANT}\\
Q_{\alpha_i,\alpha_{i+1}} & = & -
P_{\alpha_{i+1}} (1-P_{\alpha_i}) T_{2i+1} R_{2i+1}.
\label{secondCUMULANT}
\end{eqnarray}
\end{subequations}
where $P_{\alpha_i} = R_{2i}$ ($P_{\alpha_i} = T_{2i}$) 
for $\alpha_i$ being odd (even).

The next nonzero cumulant is the $N$th-order cross-correlation function
\begin{equation}
Q_{\alpha_1,\alpha_2,\cdots,\alpha_N}
=2\,{\rm sign}(\alpha_1,\alpha_2,\cdots,\alpha_N)\Gamma_{\rm BS}
\cos(\phi_{\rm tot}),
\label{NthCumulant}
\end{equation}
where the sign depends on the choice of the detectors,
${\rm sign}(\alpha_1,\alpha_2,\cdots,\alpha_N)=(-1)^{N-1+\sum_i\alpha_i}$, 
the factor 
$\Gamma_{\rm BS}=\prod_{i=1}^{2N}\sqrt{T_i R_i}$ characterizes the transmission 
of beam splitters, and the total phase accumulated around the HBT loop
is $\phi_{\rm tot} =2 \pi \Phi / \Phi_0+\sum_{i=1}^{2N} \phi_i$,
with $\Phi_0 = 2\pi\hbar/e$ being the flux quantum. Below we will use 
a notation $\{\alpha\}=\{\alpha_1,\alpha_2,\cdots,\alpha_N\}$
for an arbitrary set of $N$ detectors, so that 
$Q_{\alpha_1,\alpha_2,\cdots,\alpha_N}\equiv Q_{\{\alpha\}}$.

We note several important points. 
First, the result (\ref{NthCumulant}) holds under the usual condition
\cite{Hanbury,Samuelsson} of the ``cancellation of 
paths'', which prevents dephasing due to the energy averaging: 
The difference of the total lengths of the clockwise [see Fig.\ \ref{geometry}(a)]
and counter-clockwise [Fig.\ \ref{geometry}(b)] paths
should not exceed $\tau_Cv_F$,
where $v_F$ is Fermi velocity.
Second, 
every cumulant in Eqs.\ (\ref{firstCUMULANT}), (\ref{secondCUMULANT}),
and (\ref{NthCumulant}) has the prefactor 
$(\tau_C/2\pi\hbar)\int dE(f-f_0)^k$, where $k=1,2,N$. We consider
the zero temperature limit and set this prefactor to 1. 
Next, the sign function in the equation (\ref{NthCumulant}) has two 
contributions: the term $(-1)^{\sum_i\alpha_i}$ comes from the
detector phases, while the term $(-1)^{N-1}$ originates from
the fermionic exchange effect.
Thus, in contrast to the case of the second cumulant
(\ref{secondCUMULANT}), the overall sign of high-order cumulants
is not universal. 
Finally, only the $N$th-order cross-correlator shows oscillations as a 
function of the magnetic flux threading the HBT loop.
We stress that these oscillations appear in the FCS of electrons 
injected from $N$ uncorrelated sources, and thus they can be
regarded as a $N$-particle AB effect. 
Below we connect this effect with $N$-particle GHZ entanglement.

{\it GHZ entanglement.}---
To clarify the origin of the AB oscillations 
in the cumulant $Q_{\{\alpha\}}$ in Eq. (\ref{NthCumulant}),
we analyze the multiparticle state injected from the sources,
$| \Psi \rangle = \prod_{0<E<eV} \prod_{i=1}^{N}
c^\dagger_{4i-2} (E) |0 \rangle$.
Here, $|0 \rangle$ denotes the filled Fermi sea below $E=0$, and 
the operator $c^\dagger_{4i-2}(E)$ creates an electron 
with the energy $E$ in the source reservoir $4i-2$.
Using the scattering matrix $\hat{U}$,
we write $c^\dagger_{4i-2}=-i\sqrt{R_{2i-1}}\,a^{\dagger}_{i+1}
+\sqrt{T_{2i-1}}\,b^{\dagger}_{i}$, 
where $a^{\dagger}_i$ creates an electron moving clockwise from
the beam splitter $M_{2i-1}$ to $M_{2i}$, and 
$b^{\dagger}_i$ creates an electron moving counter-clockwise from
$M_{2i+1}$ to $M_{2i}$. Then up to the overall constant prefactor
the total state becomes
\begin{eqnarray}
| \Psi \rangle = \prod_{0<E<eV}
[\gamma_a C_a^\dagger (E) +  \gamma_b 
C_b^\dagger (E)  
+ D^\dagger(E) ] |0 \rangle,
\label{Orbital}
\end{eqnarray}
where  
$\gamma_a = \prod_{i=1}^N (-i \sqrt{R_{2i-1}})$,
and $\gamma_b = -\prod_{i=1}^N \sqrt{T_{2i-1}}$.
The operator $C_a^\dagger= \prod_{i=1}^N a^\dagger_i$
creates $N$ electrons moving clockwise [see Fig.\ \ref{geometry}(a)],
and $C_b^\dagger =\prod_{i=1}^N b^\dagger_i$
creates the $N$ electrons moving counter-clockwise [Fig.\ \ref{geometry}(b)], 
while $D^\dagger$ creates the other possible
states [Fig.\ \ref{geometry}(c-h)].

It follows from Eq. (\ref{Orbital}) that the total 
state $| \Psi \rangle$ is a product of $N$-particle states
where all $N$ particles are taken at same energy. It is 
obvious that the zero-frequency $N$th-order cumulant 
$Q_{\{\alpha\}}$ originates from the independent contribution 
of such states, more precisely, from the part
$| \psi_N\rangle =(\gamma_a C_a^\dagger +  \gamma_b 
C_b^\dagger) |0 \rangle$. Introducing orbital pseudo-spin
notations $|\!\uparrow_i\rangle=a^\dagger_i|0 \rangle$ for 
electrons moving clockwise and 
$|\!\downarrow_i\rangle=b^\dagger_i|0 \rangle$
for electrons moving counter-clockwise, we rewrite the 
relevant $N$-particle state as
\begin{equation}
|\psi_N \rangle = 
p e^{i \theta_p}
|\uparrow_1\uparrow_2 \cdots \uparrow_N \rangle + 
q e^{i \theta_q} |\downarrow_1\downarrow_2 \cdots \downarrow_N \rangle,
\label{GHZ}
\end{equation}
where 
$p^2 = 1-q^{2} = |\gamma_a|^2 / (|\gamma_a|^2 + |\gamma_b|^2)$,
$\theta_p=-\pi N/2$, and $\theta_q=\pi$.
This state is nothing but the $N$-particle GHZ-type entangled state
\cite{GHZ}. We thus arrive at the important result that in fact
GHZ entanglement is responsible for the $N$-particle AB effect, 
and that the measurement of the $N$th-order cumulant $Q_{\{\alpha\}}$
effectively postselects \cite{Shih} the $N$-particle 
entangled state.

There exists an equivalent representation of the total state 
$|\Psi \rangle$ in the wavepacket basis \cite{Levitov}, which 
obviously leads to the same physical results but allows slightly 
different interpretation that we will use below. 
In this representation, electrons fully occupy the stream of wave
packets regularly approaching, with the rate $eV/(2\pi\hbar)$,
the HBT interferometer from the sources $4i-2$, $i=1,2,\ldots,N$.
After being injected to the interferometer, the wave
packets form a $N$-particle state that contains the entangled 
state $|\psi_N \rangle$. The entangled state then moves
towards the detectors $\alpha_i=\{4i-1,4i\}$, where each electron 
accumulates the phase $\pm\phi_i$, it is rotated by the beam splitters $M_{2i}$,
$a^\dagger_i= -i\sqrt{R_{2i}}\, c^\dagger_{4i-1}
+ \sqrt{T_{2i}}\, c^\dagger_{4i}$ and  
$b^\dagger_i= \sqrt{T_{2i}}\, c^\dagger_{4i-1}
- i \sqrt{R_{2i}}\, c^\dagger_{4i}$, 
and then $|\psi_N\rangle$ is detected in one of the sets 
$\{\alpha\}=\{ \alpha_1, \alpha_2, \cdots \alpha_N \}$ of $N$
detectors.

Following Ref.\ \cite{Glauber} we introduce 
the probability ${\cal P}_{\{\alpha\}}$ of joint detection (JDP) of $N$ particles 
in the detector reservoirs defined  
as  ${\cal P}_{\{\alpha\}}\propto  \langle \Psi | 
I_{\alpha_1}I_{\alpha_2}\cdots I_{\alpha_N}
|\Psi \rangle$,
where 
$I_{\alpha_i} = (e/2\pi\hbar)\int\!\!\int dEdE'
c^\dagger_{\alpha_i}(E)c_{\alpha_i}(E')$
are the current operators in reservoirs $\alpha_i$ taken at the same time $t=0$.
Then the straightforward calculation gives 
\begin{equation}
{\cal P}_{\{\alpha\}} \propto
\prod_{i=1}^N R_{2i-1} P_{\alpha_i} +
\prod_{i=1}^N T_{2i-1} (1-P_{\alpha_i}) 
+ Q_{\{\alpha\}}
\label{JointProb}
\end{equation}
with the prefactor determined by the normalization
$\sum_{\{\alpha\}}{\cal P}_{\{\alpha\}}=1$. This result can be easily 
understood after looking closely at Fig.\ \ref{geometry}. 
The wave packets moving as shown in Fig.\ \ref{geometry}(c-h)
do not contribute to ${\cal P}_{\{\alpha\}}$, because one of the 
currents $I_{\alpha_i}$ is exactly zero. Thus, ${\cal P}_{\{\alpha\}}$ 
originates 
from the states (a) and (b), i.e., from the GHZ-type state 
(\ref{GHZ}). In equation (\ref{JointProb}) the first and the
second terms come from the direct contribution of the spin-up 
and spin-down parts of the state (\ref{GHZ}), respectively, 
while the third, $Q_{\{\alpha\}}$, 
originates from the overlap of the two parts. Thus we conclude 
that the JDP ${\cal P}_{\{\alpha\}}$ 
postselects from the Fermi sea the entangled state (\ref{GHZ}),
which via the term $Q_{\{\alpha\}}$ leads to the $N$-particle AB effect.

{\it Quantum nonlocality.}---
The GHZ-type states (\ref{GHZ}) play a special role in quantum mechanics,
because they violate generalized Bell inequalities 
\cite{Svetlichny,MABK,ZB,Chen} and thus demonstrate quantum nonlocality 
of $N$-particle states. 
The inequalities may be formulated in terms of the $N$-spin correlation 
function  
$E_N=\langle 
\sigma_{\vec{n}_1}\otimes\cdots\otimes\sigma_{\vec{n}_N} 
\rangle$,
where $\sigma_{\vec{n}_i} = \vec{n}_i \cdot \vec{\sigma}$
is the spin operator along the direction
$\vec{n}_i = (\cos \theta_i  \sin \delta_i , \sin \theta_i
 \sin \delta_i , \cos \delta_i)$ at the $i$th spin detector.
For the GHZ-type state (\ref{GHZ}) we have
\begin{eqnarray}
E_N =
g(p,q) \prod_{i=1}^N \cos \delta_i
+ 2pq
\cos \theta
\prod_{i=1}^{N} \sin \delta_i,
\label{GHZBELL}
\end{eqnarray}
where 
$g(p,q) = p^2 + (-1)^N q^2$ and
$\theta =
\theta_p - \theta_q + \sum_{i=1}^N \theta_i$.
Turning now to the generalized HBT setup, we note that the detector
beam splitters $M_{2i}$ implement the orbital pseudo-spin rotation,
while the reservoirs $4i$ and $4i-1$ detect pseudo-spins
in $z$-direction. The pseudo-spin
correlation function can now be found as 
\begin{equation}
E_N = \sum{}_{\{\alpha\}} 
(-1)^{\sum_{i=1}^N \alpha_i}
P_{\{\alpha\}},
\end{equation}
generalizing two-particle cases \cite{Samuelsson}.
After identifying $\cos\delta_i=T_{2i}-R_{2i}$ and 
$\theta_i=\phi_{2i-1}+\phi_{2i}-\pi/2$ we find that $E_N$ 
takes the {\em same} form as (\ref{GHZBELL}) with $\theta$ being 
replaced with $\phi_{\rm tot}-\pi(N-1)$.

For a particular choice of two sets of directions 
$\{\vec{n}_1,\ldots,\vec{n}_N\}$ and $\{\vec{n}_1',\ldots,\vec{n}_N'\}$,
the expectation value of the $N$-spin operator introduced via iterations,
$M_N=\frac{1}{2}(\sigma_{\vec{n}_N}+\sigma_{\vec{n}'_N})
\otimes M_{N-1}+\frac{1}{2}(\sigma_{\vec{n}_N}-\sigma_{\vec{n}'_N})
\otimes M_{N-1}'$ (with $M_N'$ obtained from $M_N$ by exchanging all 
$\vec{n}_i$ and $\vec{n}_i'$), 
may violate the Mermin-Ardehali-Belinskii-Klyshko 
(MABK) inequalities  $\langle M_N\rangle\leq 1$ \cite{MABK} that discriminate 
local variable theories and quantum mechanics. 
These inequalities generalize the well known
Clauser-Horne-Shimony-Holt inequality \cite{CHSH},
$\langle M_2\rangle=(1/2)[E(\vec{n}_1,\vec{n}_2)+E(\vec{n}_1',\vec{n}_2)
+E(\vec{n}_1,\vec{n}_2')-E(\vec{n}_1',\vec{n}_2')]\leq 1$.
The violation of more restrictive 
Svetlichny inequalities $\langle S_N\rangle\leq 2^{(N-2)/2}$
\cite{Svetlichny}, where $S_N=(1/\sqrt{2})(M_N+M_N')$ for $N$ being odd and $S_N=M_N$ 
otherwise, rules out all hybrid local-nonlocal models \cite{Gisin}. 

Here we focus on the sufficient condition for the violation of generalized
Bell inequalities, which can be found as follows: After fixing 
$T_{2i}=R_{2i}=1/2$ for all detectors the maximum values
$\langle M_N\rangle_{\rm max}=2pq2^{(N-1)/2}$ \cite{Scarani} and 
$\langle S_N\rangle_{\rm max}=2pq2^{(N-1)/2}$ \cite{Gisin} 
can be reached for a particular choices of detector phases $\phi_i$.
This will violate MABK inequalities if $2pq>1/2^{(N-1)/2}$,
and Svetlichny inequalities if $2pq>1/\sqrt{2}$.
We note that according to equations
(\ref{NthCumulant}) and (\ref{JointProb}) the value $2pq$
is nothing but the visibility $V_{\rm AB}=({\cal P}_{\{\alpha\}}^{\rm max}
-{\cal P}_{\{\alpha\}}^{\rm min})/
({\cal P}_{\{\alpha\}}^{\rm max}+{\cal P}_{\{\alpha\}}^{\rm min})$
of AB oscillations in the JDP 
for a fixed set $\{\alpha\}$
of detectors. Thus, we come to the important 
practical conclusion that the observation of sufficiently strong
AB oscillations in ${\cal P}_{\{\alpha\}}$,
\begin{equation}
V_{\rm AB}>1/\sqrt{2},
\label{condition}
\end{equation} 
will guarantee the possibility of the violation of Svetlichny inequalities
\cite{footnote2}. 
This is also true in the case of a weak dephasing, 
since in our HBT setup,
where single pseudo-spin flips are not allowed \cite{Samuelsson}, its only effect 
is to suppress the second term in the correlator (\ref{GHZBELL}).
Summarizing this discussion we conclude that the $N$-particle AB effect 
may be viewed as a manifestation of genuine quantum nonlocality in the 
generalized HBT setup.

{\it Feasibility of experimental realization.}---
The mesoscopic implementation of two-particle HBT setup 
proposed in Ref.\ \cite{Samuelsson} may be well 
utilized in the cases of $N\geq3$. It relies on the 
quantum Hall edge states as chiral channels, and quantum point 
contacts as beam splitters, and generalizes the electronic 
Mach-Zehnder interferometer, which has been recently 
experimentally realized \cite{MZ}. 
All limitations not specific to $N\geq 3$ can be found in \cite{Samuelsson}.

Recent experiments 
\cite{Reulet} 
revealed a number of specific difficulties in measuring FCS.
First of all, the detection of current fluctuations on long 
time scale $t\gg\tau_C=2\pi\hbar/|eV|$ reduces the signal-to-noise 
ratio for the $N$th-order cumulants by the factor $(\tau_C/t)^{N/2-1}$, 
the consequence of the central limit theorem. This may dramatically 
increase the total measurement time for high-order cumulants. 
Second, experimentally measurable high-order cumulants
contain nonuniversal low-order corrections from electrical 
circuit, which makes it difficult to extract intrinsic noise.  
We believe however that all these difficulties should not be that
severe in our case, because low-order cumulants do not contain AB 
oscillations and appear 
merely as a background contribution.    

Finally, in contrast to the two-particle case in Ref.\ 
\cite{Samuelsson}, for the demonstration of the quantum nonlocality 
in the generalized HBT setup the measurement on the short 
time scale $t<\tau_C$ is preferable. This is because the nonlocality condition 
(\ref{condition}) may requires a more accurate  determination of the 
visibility $V_{\rm AB}$ via measuring the JDP. Such 
high-frequency ``quantum noise'' detection techniques
have recently become available \cite{quantum}. 

We thank M.\ B\"{u}ttiker and P.\ Samuelsson for discussions,
APCTP focus program on {\it Quantum Effects in Nanosystems},
and the support from the KRF
(KRF-2005-003-C00071, C00055) and the Swiss NSF.

\bibliographystyle{apsrev}

\begin{thebibliography}{99}
\bibitem{footnote1}
Because of the chiral nature of transport in the HBT setup,
the average current defined by (\ref{Generator-A}) is divergent. 
This divergence is canceled by the outgoing noiseless current
from detectors, which is taken into account in Eq.\ (\ref{firstCUMULANT}).
\bibitem{footnote2}
MABK inequalities lead to a less restrictive condition,
$V_{\rm AB}>1/2^{(N-1)/2}$. Moreover, for $N=3$
\cite{Chen} and for $N\geq4$ being even \cite{ZB}, they
can be generalized and violated for arbitrary weak visibility,
ruling out all local variable theories. However, none of these 
discriminate hybrid local-nonlocal theories
and quantum mechanics.
\end{thebibliography}

\end{document}